\documentclass[aps,twocolumn,showpacs,prl,amsfonts]{revtex4}
\usepackage{epsfig,amsmath}
\newcommand{\Ref}[1]{(\ref{#1})}
\newcommand{\REF}[1]{Eq.~(\ref{#1})}
\def \BSE {\begin{subequations}}
\def \ESE {\end{subequations}}
\def \BE {\begin{equation}}
\def \EE {\end{equation}}
\def \BEA {\begin{eqnarray}}
\def \EEA {\end{eqnarray}}

\newcommand{\p}{\partial}           

\def\<{\left \langle} \def\>{\right\rangle}

\begin{document}
\title{Differential  model for 2D turbulence}
\author{
Victor S. L'vov$^{*\ddag}$ and Sergey Nazarenko$^\dag$}
 \affiliation {
$^*$ Department of Chemical Physics, the
Weizmann Institute of Science, Rehovot  76100, Israel \\
$^\dag$   Mathematics Institute, The University of Warwick,
Coventry, CV4-7AL, UK\\
 $^\ddag$ Low Temperature Laboratory, Helsinki University of Technology,
P.O. Box 2200, FIN-02015 HUT, Finland }

\date{\today}

\begin{abstract}
We present a phenomenological model for 2D turbulence in
which the energy spectrum obeys a nonlinear fourth-order
differential equation. This equation respects the
scaling properties of the original Navier-Stokes
equations and it has both the $-5/3$ inverse-cascade and
the $-3$ direct-cascade spectra. In addition,our model
has  Raleigh-Jeans thermodynamic distributions, as exact
steady state solutions. We use the model to derive a
relation between the direct-cascade and the
inverse-cascade Kolmogorov constants which is in a good
qualitative agreement with the laboratory and numerical
experiments. We discuss a steady state solution where
both the enstrophy and the energy cascades are present
simultaneously and we discuss it in  context of the
Nastrom-Gage spectrum observed in atmospheric
turbulence. We also consider the effect of the bottom
friction onto the cascade solutions, and show that it
leads to an additional decrease and finite-wavenumber
cutoffs of the respective cascade spectra which agrees
with existing experimental and numerical results.
\end{abstract}
\maketitle

\subsection{I. Leith differential model and its relatives} In 1967 paper
\cite{leith}, Leith proposed the following differential equation
model as a model of 3D turbulence,
\begin{equation}
\frac{\partial E}{\p t} = {1 \over 8}\, \frac{\p}{\p
k}\Big[\sqrt{k^{11}\, E }\, \frac{\p}{\p
k}\Big(\frac{E}{k^2}\Big)\Big] - \nu k^2 E, \label{fp}
\end{equation}
where $t$ is time, $k$ is the absolute value of the wavenumber,
$\nu$ is the kinematic viscosity and the one-dimensional energy
spectrum $E(k,t)$ is normalised so that the kinetic energy density
is $\int E \, dk$. The third term on the RHS of Eq.~(\ref{fp}) is
obvious and describes the linear process of viscous dissipation.
The first term contains a nontrivial model of the nonlinear
processes in turbulence which rests on three basic properties:

1.  The total energy density  $\int E \, dk$ is conserved by the
inviscid dynamics. The characteristic time of the spectral energy
redistribution for {\em local} interaction of turbulent scales is of
order of the vortex turnover time, $1/\sqrt{k^3 E}$.

2. The steady state in forced turbulence corresponds to a constant
energy cascade though the inertial range of scales which is
described by the Kolmogorov spectrum,
\begin{equation}
E = C \, P^{2/3} \, k^{-5/3},
\label{kolm}
\end{equation}
where $P$ is the energy flux (constant in $t$ and $k$) and $C \approx 1.6$ is
the Kolmogorov constant.

3. When the wave-number range is truncated to a finite value and
both forcing and dissipation are absent,  turbulence reaches a
thermodynamic equilibrium state characterized by equipartition of
energy over the wave-number space \cite{kraichnan}.
 In terms of the one-dimensional energy spectrum this means
$E \propto k^2$ which is obviously a steady state solution of the
equation (\ref{fp}) for $\nu =0$.

Even simpler first order differential equation was proposed by
Kovazhny, which can model the cascade properties but ignores the
thermodynamic distributions. Due to their simplicity, Leith and
Kovazhny models and their generalisations to include other physical
effects turned out to be very efficient. In particular, they were
used for description mixed cascade-thermo solutions corresponding to
the bottleneck effect \cite{cn}, superfluid turbulence in $^3$He
with friction against the normal component \cite{lnv}, influence of
heavy particles on turbulent suspensions \cite{lvov1}.

Differential models have also been proposed and used before for wave
turbulence. In particular, Hasselmann and Hasselmann \cite{hass}
and, independently, Iroshnkov \cite{irosh} (thereafter HHI) proposed
a model for the spectrum of deep-water surface gravity waves, which
is a fourth-order nonlinear differential equation. The order of this
equation is higher than in the Leith model because, in addition to
the direct energy cascade and the thermodynamic energy
equipartition, it describes the inverse cascade and the
equipartition state of waveaction, - an additional invariant of the
water wave system. Recently this model was used to describe a
sandpile-like behaviour energy cascades of waves in discrete
$k$-space \cite{sand}. Similar model was proposed for turbulence of
Kelvin waves on thin quantum vortices, and it was used to study
sound radiation by such waves \cite{kelvin}. This model is also of
the fourth order because the Kelvin wave system also conserves
waveaction~\footnote{ Waveaction is an adiabatic invariant
(approximately) conserved by weakly nonlinear systems whose resonant
interactions are of even order. For example, on the deep water the
leading resonances are four-wave whereas for the Kelvin waves they
are six-wave.}.

\subsection{II. Differential Model for 2D turbulence}

A 2nd-order differential model for 2D turbulence was proposed by
Lilly \cite{lilly2}, see equation (\ref{2ndorderDA}) below. It has
both the energy and the enstrophy cascade solutions, but it ignores
the thermodynamic equilibria. As we will see later, this causes a
significant defect in description because it fails to predict the
important difference in values of the Kolmogorov constants for the
direct and the inverse cascade spectra. The minimal model that would
include both cascades and the thermodynamic solutions has to be of
the fourth order. In some sense, such model should be a hybrid of
the Leith model, since it has to respect the Navier-Stokes scalings,
and the HHI model, because of presence of the additional invariant -
enstrophy. We therefore postulate the following 4th-order evolution
equation for the energy spectrum in 2D turbulence,
\begin{equation}
\frac{\p E}{\p t} = a \sqrt{x}\frac{\p^2 }{\p x^2} \Big[ \sqrt{
E^5 x^{19/2 -D }} \frac{\p^2 }{\p x^2} \Big(\frac{\sqrt{x^{D -1}}}
{ E} \Big) \; \Big] - \nu x E, \label{dm2d}
\end{equation}
where $x=k^2$ and $D$ is the dimension of the system. Although our
main object is 2D turbulence, $D=2$, we would like to keep $D$ as
a parameter for the purposed which will be clear later. For
$\nu=0$ this equation conserves the mean energy density, 
\BSE
\begin{equation}
{1 \over 2} \langle v^2 \rangle = \int E \, dk = \int {E \, dx
\over 2 \sqrt{\,x\,}}\,, \label{energy}
\end{equation}
and the mean enstrophy density,
\begin{equation}
{1 \over 2} \langle (\nabla \times v)^2 \rangle = \int k^2 E \, dk
= {1 \over 2} \int {\sqrt{\,x\,}}  E \, dx\ . \label{enstrophy}
\end{equation}
\ESE
 Equation (\ref{dm2d}) can be written as a continuity
equations for the energy, $E$, and  the enstrophy, $\Pi = k^2 E$,
spectra: 
  \BE   \frac{\p E}{\p t } + \frac{\p P}{\p k } = 0\,, \quad
\frac{\p P}{\p t } + \frac{\p Q}{\p k } = 0\, , \label{cont2} \EE
  where $P$ and $Q$ are the flux of energy and the flux of
enstrophy respectively,
 \BSE\label{flux1}
 \BEA
 P &=& -{a \over 2} \,\frac{\,\partial R\,}{\p x }\,, \quad
Q =  -{a \over 2} \Big(R - x \,\frac{\,\partial R\,}{\p x } \Big)\,,
\label{q}\\
 R &=& \sqrt{  E^{\,5}\, x^{19/2 -D}}
\,\frac{\,\partial^2 \,}{\p x^2 } \Big({ \sqrt{x^{D -1}} \over E}
\Big)\ . \label{R} \EEA\ESE

\subsection{III. Steady state solutions}
Equation (\ref{dm2d}) has a general thermodynamic Raleigh-Jeans
solution in which both the energy flux and the enstropy flux are
zero. It is determined by the condition $R=0$ which yields,
\begin{equation}
E = (T \, k^{D -1})/  (k^2 + \mu)\, , \label{rj}
\end{equation}
where constants $T$ and $\mu$ are    temperature and chemical
potential respectively.

In addition, Eq.~(\ref{dm2d}) has two cascade solutions: an inverse
energy cascade solution ($P= $ const.), \BSE\label{flux}
\begin{equation} E = C_P \, (-P)^{2/3} \, k^{-5/3}\, ,
\label{invcascade}
\end{equation}
and a forward enstrophy cascade state ($Q=$ const.),
\begin{equation}
E = C_Q \, Q^{2/3} \, k^{-3},
\label{forwcascade}
\end{equation}\ESE
where $C_P$ and $C_Q$ are  Kolmogorov constants.

Substituting Eqs. \Ref{flux}  into Eqs.~\Ref{flux1}, one can find
relations between $C_P$, $C_Q$ and   $a$:
\BSE\BEA
 72  &=& a\,    C_P \,  (3D+  2)\Big(3D -  4
  )\, , \label{a_P}\\
 8 &=&  a \, C_Q \, {D }(D + {2})\ .
\label{a_Q} \EEA
 This implies the following relation between the
Kolmogorov constants,
\begin{equation}
  {C_Q \over  C_P}   =
\left[{ (3D+2)(3 D - 4) \over {9\, D }(D + 2)}\right]^{2/3}\ .
\label{P_Q}
\end{equation}\ESE
Note that this expression turns into zero for $D=4/3$ which means
that the the for this dimension the inverse cascade rate turns into
zero and the corresponding inverse-cascade spectrum degenerates into
thermodynamic energy equipartition. Correspondingly, one can expect
that for $D=4/3$ Gaussian statistics for this range of scales
(up-scale from the forcing scale).

For $D=2$ we have
\begin{equation} {C_Q \over  C_P}  =
\left( { 2 \over9} \right)^{2/3} \approx 0.367.
\label{P_Q_2D}
\end{equation}
 Note that $D=2$ is rather close to the critical dimension $D=4/3$
and, therefore, statistics of turbulence in the inverse-cascade
range should be expected to be quite close to Gaussian. This
observation was previously made in \cite{lvov2}. Another
consequence of the fact that $D=2$ is close to the critical
dimension $D=4/3$ is that ${C_Q } $ is significantly less than ${
C_P}$ as seen in (\ref{P_Q_2D}). Experimental values for ${  C_Q}$
are $1.4 \pm 0.3$ \cite{tabeling} where as DNS give $1.6 \pm 0.1$
\cite{borue} and more recently $1.9 \pm 0.1$ \cite{kaneda}.
 For  ${  C_P}$, experiments give ${  C_P} \approx
6.5 \pm 1$ \cite{tabeling2} which is consistent with DNS (e.g.
$C_P \approx 7$ in \cite{smith_yakh}).
More recently it was argued however \cite{gurarie} that  the energy
condensation at largest scales makes it difficult to measure
$C_P$ and its fit varies from $4.5$ near the forcing scale
to about $7.2$ near the peak of the spectrum.
Taking a ratio
corresponding to most recent high-resolution
DNS we have $C_Q/C_P \approx 1.9/6
\approx 0.32$ which is in a godd qualitative agreement with
(\ref{P_Q_2D}). Note that although there is still a significant
spread in the available experimenta and numerical data on the
Kolmogorov constants, their ratio is always significantly
less than one and our model predicts and explains this behavior.

\subsection{IV. Two-flux solutions}

Pure single cascade states \Ref{flux}  are mathematical
idealisations corresponding to infinitely wide inertial ranges,
whereas in both numerical simulations and nature these ranges are
finite and typically both the energy and the enstrophy cascades are
present in important turbulent scales. In particular, such
two-cascade states are often applied for explaining the Nastrom-Gage
spectrum observed in atmospheric turbulence: with $-3$ exponent at
low wavenumbers and $-5/3$ at the high wavenumber end
\cite{ng,ng1,lilly2}. To find a analytical expression for such
two-flux spectra, Lilly introduced a second-order equation model
\cite{lilly2} which ignores the thermodynamic states but describes
both the inverse and the forward cascade states,
\begin{equation}
\frac{\p E}{\p t} = b \, \frac{\p }{\p k}\,\Big[ \sqrt{\, k\, E\,}
\, \frac{\p }{\p k}\left(k^3 E\right) \Big] - \nu k^2 E,
\label{2ndorderDA}
\end{equation}
where $b$ is a dimensionless constant which has to be found
from an experimental or numerical value of $C_P$ (using $C_Q$
is less desirable due to certain lack of precision in
finding this constant related to nonlocality and respective
log-correction of the direct cascade state). We have,
\begin{equation}
b = 3 / ( 4 \, C_P^{3/2} )\approx 0.05\ . \label{b}
\end{equation}

Equation (\ref{2ndorderDA}) is written in a form on the energy
continuity equation. It can also be re-written as an enstrophy
continuity equation,
\begin{equation}
\frac{\p \,(k^2 E) }{\p t}  = b\,  \frac{\p }{\p k}\Big[ \sqrt{
k^{23/3} \, E\, } \frac{\p }{\p k} \left( k^{5/3} E\right) \Big] -
\nu k^4 E\ . \label{2ndorderDAens}
\end{equation}
Such a 2nd-order model is less realistic than the 4th-order model
because it gives $C_P=C_Q$ due to absence
 of thermodynamic states. As we will see below, this also makes it
less realistic for description of the two-flux states.

The general steady-state solution of (\ref{2ndorderDA}) is
\begin{equation}
E = k^{-3} \left(C_P^{3/2} |P| k^2 + C_Q^{3/2} Q\right)^{2/3}\ .
\label{mixed}
\end{equation}
This spectrum exhibits behaviour characteristic of the Nastrom-Gage
spectrum: $-3$ exponent at the left and $-5/3$ right sides of the
inertial range \cite{ng,ng1}. Formally, $C_P=C_Q$, and substituting
this equality into (\ref{mixed}) we would recover a solution
originally obtained by Lilly \cite{lilly2} (whose main goal was to
explain the Nastrom-Gage spectrum). However, we emphasize that the
more precise 4th-order  model, as well as the DNS and experimental
data, give significantly different values for $C_P$ and $C_Q$, and
therefore we can expect that Eq.~(\ref{mixed}) to work better with
$C_P$ and $C_Q$ taken from the 4th-order model, or from the
available numerical or experimental data discussed above.
 Transition between the
two exponents occurs at \BE 
k_* \sim \sqrt{\,C_Q^{3/2} Q \left / C_P^{3/2} |P|\right . \, }\
.
\label{k_star}
\EE
As it was pointed out in \cite{lilly2}, this transition occurs
without any   sinks of turbulence in between of the two sources.

Formula (\ref{k_star})   gives correct asymptotics for $P \gg Q$ and
for $P \ll Q$. However, by substituting it into more precise
fourth-order model we see that both the energy and the enstropy
fluxes (\ref{q}) experience order of magnitude variation in the
region $P \sim Q$  instead of remaining approximately constant as
they should. Thus, we conclude that (\ref{k_star}) does not work
well in the transitional region $P \sim Q$ and a better description
should be expected from finding the two-flux states directly in the
4th-order model (\ref{dm2d}). Using Eqs.~\Ref{flux1}, we can write
the following differential equation for the general two-flux steady
state,
\begin{equation}
\sqrt{  E^{\,5}\, x^{15/2}} \,\frac{\,\partial^2 \,}{\p
x^2 } \Big({ \sqrt{x} \over E} \Big) = -2(Px +Q)/a \ .
\label{2f4}
\end{equation}
This equation cannot be solved analytically, but it can be easily
integrated numerically for any particular values of $P$ and $Q$ and
compared to observational or DNS data.

Note, however, that the two-source scenario is not the only suggested
explaination of  Nastrom-Gage spectrum and, in particular, the steeper slope
at small wavenumbers can appear due to non-universality because of
ineficiency of large-scale dissipation resulting in condensation
of energy at the large scales (see detailed discussion and references in review
\cite{danilov}).

\subsection{V. Effect of friction}

In  many situations, particularly atmospheric turbulence and laboratory
experiments, the
bottom friction is believed to have an important effect on 2D
turbulence, and in this section we will address this issue using
our differential model. Our interest here is partially motivated
by a question if the bottom friction can help to form a spectrum
of the Nastrom-Gage shape, particularly in the view of results of
\cite{lnv} where a spectrum with transition from the $-3$ to the
$-5/3$ exponent was obtained in 3D superfluid turbulence with
friction (and see comments at \footnote{Similar question was
examined by Elef Gkioulekas in his talk at the Warwick Symposium
workshop `` Universal features in turbulence: from quantum to
cosmological scales''
 (December 5-10, 2005). However he used a 3D differential model
which is less relevant for this case than the 2D model introduced
in the present paper and therefore we re-examine this question
using our model.}).  For all that follows one could use the fourth
order equation, but we chose to work with the second order model
here because of easier algebra. Adding friction and ignoring
viscosity in (\ref{2ndorderDA}) we have,
\begin{equation}
\frac{\p E}{\p t} = b \, \frac{\p }{\p k}\,\Big[ \sqrt{\, k\, E\,}
\, \frac{\p }{\p k}\left(k^3 E\right) \Big] - \gamma E\ ,
\label{2ndorderDAf}
\end{equation}
where $\gamma$ is a $k$-independent  friction frequency. One could
try a power law substitution, $E = C k^y$, for which the power
counting immediately gives $y=-3$ \cite{smith_yakh2},  but the
constant $C$ turns out to be undefined (infinite). This is natural
because the $-3$ shape coincides with the direct cascade spectrum in
the absence of friction and, thus, the nonlinear term in
(\ref{2ndorderDAf}) turns into zero. However, there should be a
corrected solution which corresponds to a direct enstrophy cascade
in presence of friction. Close to the forcing scale $k_F$ this
correction can be found substituting $E = C_Q \, Q^{2/3} \, k^{-3}(1
+ \delta_Q)$
 into (\ref{2ndorderDAf}) which gives after linearisation,
\begin{equation}
\delta_Q = - {\gamma \over 2b}  C_Q^{1/2} \, Q^{1/3} \ln (k/k_F)\ .
\label{deltaQ}
\end{equation}
One can see that this correction is negative and growing in
magnitude
 along the cascade to high $k$'s.
Far from the forcing scale, $k \gg k_F$ we gain
a logarithmic factor,
\begin{equation}
E =  {\gamma^2 C_Q^3 \over 9\, k^3} \,   \Big[\ln \Big(\frac{k_Q}{
 k}\Big) \Big]^2\, , \label{log}
\end{equation}
where $k_Q$ is a constant wavenumber at which an abrupt cutoff of
the spectrum occurs. Note that the finite cutoff is a result of
using the differential model and it indicates that in more general
models the spectrum, although always non-zero,  decays at these
scales faster than any power of $k$. The value of $k_Q$ can be
obtained via numerical solution for the full spectrum matching the
regions close and far from the forcing scale. Note that because
neither $\gamma$ nor $Q$ contain the length dimension the expression
for $k_Q$ must contain the forcing wavenumber - the only relevant
quantity containing the length dimension.

Result (\ref{log}) about the log-correction is natural keeping in
mind the degeneracy of coincidence of the direct cascade exponent,
$-3$, with the exponent arising from dimensional balancing of the
friction and the inertial terms in the energy balance equation.
However, papers \cite{bof1,bof2} present a numerical and
experimental evidence that the friction effect on the direct cascade
is to modify the spectrum exponent rather that to log-correct it.
Plausible explanations for the descrepancy with the prediction of
the differential model could be: (i) energy condensation and
nonlocal interactions with large scales may have been important in
\cite{bof1,bof2} (they are not described by the differential model
which is super-local) or (ii) insufficient fitting interval (one
decade) in \cite{bof1,bof2} did not allow to distinguish between the
log and power corrections. Thus, additional numerical experiments at
higher resolution are desirable to obtain a larger fitting interval.

Summarising results of (\ref{deltaQ}) and (\ref{log}), we conclude
that friction leads to a faster decay of the inverse cascade spectrum
and its eventual abrupt cutoff. This means in particular that it
would never take a flatter -5/3 slope at high $k$ characteristic
for  Nastrom-Gage spectrum.

Let us now find the friction correction to the inverse cascade
spectrum close to the forcing scale by substituting
$E = C_P \, P^{2/3} \, k^{-5/3}(1 + \delta_P)$
 into (\ref{2ndorderDAens}) [with added friction term
as in (\ref{2ndorderDAf})]. After linearisation, we have
\begin{equation}
\delta_P = - {9 \gamma \over 8b} \, C_P^{1/2} \, P^{1/3} k^{-2/3} \
. \label{deltaP}
\end{equation}
One can see that this correction is negative and growing in
magnitude
 along the cascade to low $k$'s.
Easy to see by inspection that the inverse cascade spectrum also
has a finite cutoff close to which we have
\begin{equation}
E = {\gamma^2 \over 4 b^2} (k - k_P)^2.
\label{cutinv}
\end{equation}
Thus we see that in the inverse cascade range one could never have
slope $-3$ at low $k$ characteristic for  Nastrom-Gage spectrum.
Exact position of the cutoff can be found only   by matching the
large-scale and the small-scale parts of the solution, which can be
done numerically. However, up to an order-one factor, it can be
uniquely found from the dimensional argument which gives,
\begin{equation}
k_P \sim \lambda^{3/2}/|P|^{1/2}.
\label{kq}
\end{equation}
In fact, existence of such an abrupt cutoff was suggested
  by Lilly \cite{lilly1} that the Kolmogorov relation between
the spectrum and the energy flux persist. This assumption is valid
only when friction affects the flux only weakly, and therefore it
is natural that Lilly's expression for the spectrum agrees with ours
near the forcing scale (\ref{deltaP}) but it fails to predict correct
behaviour near the cutoff scale (\ref{cutinv}). Estimate
(\ref{kq}) was also obtained by Manin who considered stability of
large-scale vortices \cite{manin}. This cutoff scale was also
discussed in experimental works \cite{dolzh,cdv,som}
as well as in numerical works (\cite{danilov} and
references therein) in the context
of arrest of the inverse energy cascade by bottom friction.

Combining the results for the direct and the inverse cascade
spectra, we conclude that when forcing is present at only one scale
the Nastrom-Gage spectrum could not be explained by the friction
effect neither in the direct cascade nor in the inverse cascade
ranges. Thus, 2D turbulence with friction is fundamentally different
from the 3D case, and the origin of this difference can be
attributed to presence of the enstrophy cascade (with exponent
coinciding with the one obtained by the formal power counting in the
in the frictional system) and to the fact that the energy cascades
in 2D and 3D are in the opposite directions with respect to each
other.~\\

 { \bf Summary.~~} In this Letter, we presented a differential models
 for 2D turbulence based on the 4th-order Eq.~(\ref{dm2d}).
 Based on \REF{dm2d} we derived the ratio of the Kolmogorov constants 0.37 which
qualitatively agrees with an experimentally and numerically measured
value of about 0.32. Of particular importance is the fact that the
model predicts the inverse-cascade constant to be significantly
greater than the direct-cascade constant. Based on a reduced
2nd-order model \Ref{2ndorderDA}, we discussed a mixed two-cascade
solution~(\ref{mixed}) which was previously suggested by Lilly for
explaining the the observed Nastrom-Gage spectrum of atmospheric
turbulence. We showed that, inspite of giving correct asymptotic
expressions in the small and large $k$ limits, this solution is
inacurate in the intermediate range. We concluded that the 4th-order
model should be used instead for finding the two-cascade states, and
doing so leads to the 2nd-order nonlinear ODE (\ref{2f4}) which can
be solved numerically in each particular case with given values of
$P$ and $Q$.  We also examined the effect of friction on the inverse
and direct cascades, and showed that in the case with only one
forcing region present, the friction cannot lead to the Nastrom-Gage
shape. Instead, for both cascades friction acts to reduce the
spectra and to terminate them abruptly at finite cutoff wavenumbers,
which agrees with previous numerical and experimental studies.
However,   our result that friction acts to introduce a
log-correction on the direct cascade rather that to modify its
exponent is different from the conclusions of previous studies
\cite{bof1,bof2}. A possible explanation of this descrepance is that
non-local effects were important in simulations and experiments of
\cite{bof1,bof2}.
\\~\\

{\bf Acknowledgements.}~~
 This work was done as the result of our
collaboration during the the Warwick Symposium workshop
``Environmental Turbulence: from Clouds through the Ocean" at
Coventry, UK on March 2006,  and the ULTI Users Meeting at
Lammi, Finland on April 2006. We are acknowledge the support of
this project by
Large Scale Installation Program ULTI of
the European Union (contract number: RITA-CT-2003-505313), EPSRC
Turbulence Symposium grant,
 the European Science Foundation Program COSLAB and
the US-Israel Binational Science Foundation.

\end{document}